\documentclass[11pt,a4paper]{article}
\usepackage{amsmath}
\usepackage{amssymb}
\usepackage{graphicx}
\usepackage{appendix}

\input{tcilatex}
\begin{document}

\title{Description of Nonlinear Phenomena in the Atmospheric Dynamics
through Linear Wave type Equations}
\author{Rodica Cimpoiasu, Radu Constantinescu \\
University of Craiova, 13 A.I.Cuza, 200585 Craiova, Romania}
\date{}
\maketitle

\begin{abstract}
The paper takles with a procedure which allow to extend some linear, wave
type models currently used in describing phenomena appearing in atmosphere
to the study of nonlinear models. More concretely, we present a practical
way to generate the largest class of $(1+1)$-dimensional second order
partial differential equations (pdes) of a given form which could be reduced
to an imposed ordinary wave type equation. This class generalize the
ordinary differential equation describing the equatorial trapped waves
generated in a continuously stratified ocean and will be obtained following
the Lie symmetry and similarity reduction procedures. Moreover, some
concrete nonlinear second order differential equations will be proposed as
possible candidates for replacing more complicated, nonintegrable systems,
as the Rossby type equation.

\textbf{Keywords}\textit{: }Nonlinear dynamical systems,\textit{\ }Lie
symmetries, Similarity reduction procedure, Rossby type symmetries.
\end{abstract}

\section{ Introduction}

A rich variety of complex phenomena occuring in many physical fields,
including the atmospheric dynamics, are described by \textit{linear
differential equations} which allow a simple handling of the constraints
which appear in the system's evolution. Although, the linearized models
often do not adequately describe the dynamics of the processes as a whole
and it is very simple to shift the system to a region in which the linear
behavior is no longer valid. This is why, in order to capture the real
behavior, to accurately estimate and control the complex systems in all
their regimes, \textit{nonlinear models} must be defined. In this case, the
linear differential equations could appear as approximations to the
nonlinear systems, valid under restricted conditions.

The price to be paid when nonlinearity is taken into consideration appears
in the investigation of the exact solutions of the attached equations. There
are not standard methods of solving nonlinear differential equations, they
are usually depend on the form of the equations and on their particular
symmetries. Many interesting nonlinear models have been proposed over the
last years \cite{Poly} and a lot of methods have been developed in order to
find solutions of equations describing these nonlinear phenomena. Some of
the most important methods \cite{Liu} are the inverse scattering method \cite%
{Gardner}, the Darboux and B\"{a}cklund transformations \cite{Rogers}, the
Hirota bilinear method \cite{Wazwaz}, the Lie symmetry analysis \cite%
{Olver,Bluman,Cantwell}, etc. By applying these methods, many types of
specific solutions have been obtained. For example, solitary waves or
solitons, which\textbf{\ }have no analogue for linear partial differential
equations, are very important for the nonlinear dynamical systems.

In this paper we shall concentrate our attention to the \textit{Lie group
method}. It is well-known that this method is a powerful and direct approach
to construct many types of exact solutions of nonlinear differential
equations, such as soliton solutions, power series solutions, fundamental
solutions \cite{Craddock,Lennox}, and so on. The existence of the operators
associated with the Lie group of infinitesimal transformations allows the
reduction of equations to simpler ones. The similarity reduction method for
example is an important way of transforming a $(1+1)-$dimensional pde into
an ordinary differential one. We shall concretely consider \textit{the
inverse symmetry problem }\cite{Cimpoiasu} and we shall generate the largest
class of second order $(1+1)-$dimensional pdes which generalize the
ordinary, wave type, differential equation describing the equatorial trapped
waves generated in a continuously stratified ocean. Practically, four types
of waves appears in this case and have to be found among the solutions of
the equation: Kelvin waves, Rossby waves, inertia-gravity waves and mixed
Rossby-gravity waves. In the Boussinesq approximation and on an equatorial $%
\beta $-plane, the equation which describe the $m$-th oscillation mode of
the wave's vertical velocity $\phi _{m}(z)$ on the direction $z$ has the
form \cite{Rossby2}:%
\begin{equation}
\frac{\partial ^{2}\phi _{m}(z)}{\partial z^{2}}+\frac{N_{m}^{2}(z)}{%
C_{m}^{2}}\phi _{m}(z)=0  \label{0.1}
\end{equation}%
One consider for the velocity the boundary conditions:%
\begin{eqnarray*}
\phi _{m}(z &=&-H)=0\text{ (ocean floor)} \\
\phi _{m}(z &=&0)=0\text{ (ocean surface)}
\end{eqnarray*}%
In the equation (\ref{0.1}), $C_{m}$ is a constant and $N_{m}(z)$ represents
the "buoyancy" frequency. Measurements made during El Ni\~{n}o events \cite%
{Rossby2} show that the buoyancy frequency $N_{m}(z)$ has strong variations
with the water depth close to the surface and practically vanishes for
higher depths. We notice that in the first case (at the surface, $z\in
\lbrack 0,300]~m$) one can aproximate $N_{m}(z)$ with an averaged value
around $\overline{N(z)}=2\cdot 10^{-4}m\cdot s^{-1}$ So, the equation (\ref%
{0.1}) can be linearized in one of the following forms:%
\begin{equation}
\overset{\cdot \cdot }{\phi }(z)=0;z\geq 300  \label{0.2}
\end{equation}%
\begin{equation}
\overset{\cdot \cdot }{\phi }(z)+k^{2}\phi (z)=0;z\in \lbrack 0,300];k\equiv 
\frac{N}{C}=const.  \label{0.3}
\end{equation}

In this paper, we shall consider the two previous wave type equations and we
shall see how they can be extended towards $(1+1)-$second order differential
equations with the same group of symmetries as the initial ordinary wave
type equations have.

The outline of this paper is as follows: after this introductory notes, in
Section 2, we shall obtain the general determining system for a chosen class
of $(1+1)-$dimensional models. The system will be generated by using the Lie
symmetry approach and by asking for an imposed form of the similarity
reduction equation. More exactly, we shall generate a class of $(1+1)-$%
second order differential equations which by similarity reduction come to
the wave forms (\ref{0.2}) and, respectively, (\ref{0.3}). The general
results of the second section will be particularized in Section 3, when\
concrete examples of two dimensional equations with similar solutions as the
ordinary wave equations (\ref{0.2}) and (\ref{0.3}) will be generated.
Moreover, we shall compute the form of the second order partial differential
equation which admit an imposed form of symmetry, specific for the two
dimensional Rossby type equation. So we shall be able to replace the study
of this last strongly nonintegrable equation with a simpler class of
equations observing similar symmetries. Some concluding remarks will end the
paper.

\section{Determining equations for the Lie symmetry group}

Let us consider the class of general dynamical systems described in a $(1+1)$%
-dimensional space $(x,t)$ by a second order partial differential equation
of the form:%
\begin{equation}
u_{t}=A(x,t)u_{2x}+B(x,t)u_{x}+C(x,t)u\Leftrightarrow \Omega
(x,t,u,u_{x},u_{t},u_{2x})=0  \label{1.1}
\end{equation}

Our aim is to select from the general dynamical systems described by (\ref%
{1.1}) the class of differential equations which admit a similarity
reduction to wave type equations of the form (\ref{0.2}) and (\ref{0.3}). As
feed-back, the solution of the wave equations will be used in order to
obtain a solution for (\ref{1.1}). The procedure that will be followed
firstly implies to obtain the system of determining equations for the Lie
symmetry group of (\ref{1.1}). Then, an additional system of partial
differential equations will be generated by imposing that (\ref{1.1})
possess a reduced similarity equation of the wave type. Finally, this last
system and the Lie determining equations will be solved and the coefficient
functions $A(x,t)$, $B(x,t)$, $C(x,t)$ will be obtained.

In this section we shall apply the Lie symmetry approach for the equation (%
\ref{1.1}). Let us consider a one-parameter Lie group of infinitesimal
transformations:%
\begin{equation}
\bar{x}=x+\varepsilon \xi (t,x,u),\text{ }\bar{t}=t+\varepsilon \varphi
(t,x,u),\text{ }\bar{u}=u+\varepsilon \eta (t,x,u)  \label{1.2}
\end{equation}%
with a small parameter $\varepsilon \ll 1.$ The Lie symmetry operator
associated with the above group of transformations can be written as:%
\begin{equation}
U(x,t,u)=\varphi (x,t,u)\frac{\partial }{\partial t}+\xi (x,t,u)\frac{%
\partial }{\partial x}+\eta (x,t,u)\frac{\partial }{\partial u}  \label{1.3}
\end{equation}%
The second order equation $\Omega (x,t,u,u_{x},u_{t},u_{2x})=0$ of the form (%
\ref{1.1}) is invariant under the action of the operator (\ref{1.3}) if and
only if the following condition \cite{Olver} is verified:%
\begin{equation}
U^{(2)}(\Omega )\mid _{\Omega =0}=0  \label{1.4}
\end{equation}%
where $U^{(2)}$ is the second extension of the generator (\ref{1.3}). A
concrete computation shows that the coefficient functions from (\ref{1.1})
and (\ref{1.3}), $A(x,t),B(x,t),C(x,t),\varphi (x,t,u),\xi (x,t,u),\eta
(x,t,u),$ must satisfy the equation:%
\begin{equation}
(\varphi A_{t}+\xi A_{x})u_{2x}+(\varphi B_{t}+\xi B_{x})u_{x}+\varphi
C_{t}u+\xi C_{x}u+C\eta +B\eta ^{x}-\eta ^{t}+A\eta ^{2x}=0  \label{1.5}
\end{equation}%
The coefficient functions $\eta ^{x},$ $\eta ^{t},$ $\eta ^{2x}$ appear in
the process of extension of $U$ towards $U^{(2)}$ and their general
expressions are given in \cite{Olver}. Using these expressions in (\ref{1.5}%
) and asking for the vanishing of the coefficients of each monomial in the
derivatives of $u(t,x)$, we obtain the following differential system:%
\begin{equation*}
\varphi _{x}=0;\ \varphi _{u}=0;\ \xi _{u}=0;\ \eta _{2u}=0;\text{ }\varphi
A_{t}+\xi A_{x}+A\varphi _{t}-2A\xi _{x}=0;
\end{equation*}%
\begin{equation}
-\varphi B_{t}-\xi B_{x}+B\xi _{x}-\xi _{t}-B\varphi _{t}-2A\eta _{xu}+A\xi
_{2x}=0  \label{1.6}
\end{equation}%
\begin{equation*}
-\varphi C_{t}u-\xi C_{x}u-C\eta -B\eta _{x}+\eta _{t}+C\eta _{u}u-\varphi
_{t}Cu-A\eta _{2x}=0
\end{equation*}%
The first four equations of the system (\ref{1.6}) lead, for coefficient\
functions $\varphi (x,t,u),\xi (x,t,u),\eta (x,t,u),$ to the following
reduced dependences:%
\begin{equation}
\varphi =\varphi (t),\text{ }\xi =\xi (x,t),\text{ }\eta =M(x,t)u\text{ }
\label{1.7}
\end{equation}%
Consequently, the remaining equations of (\ref{1.6}) become:%
\begin{equation*}
\text{ }\varphi A_{t}+\xi A_{x}+A\varphi _{t}-2A\xi _{x}=0;
\end{equation*}%
\begin{equation}
-\varphi B_{t}-\xi B_{x}+B\xi _{x}-\xi _{t}-B\varphi _{t}-2AM_{x}+A\xi
_{2x}=0  \label{1.8}
\end{equation}

\begin{equation*}
-\varphi C_{t}-\xi C_{x}-BM_{x}+M_{t}-\varphi _{t}C-AM_{2x}=0
\end{equation*}%
with $6$ unknown functions: $A(x,t),B(x,t),C(x,t)$ provided by the
evolutionary equation (\ref{1.1}) and $\varphi (t),$ $\xi (x,t),$ $\eta
(x,t,u)$ introduced by the symmetry group of transformations (\ref{1.2}) and
described by the relations (\ref{1.7}).

\section{Similarity reduction procedure}

Let us consider now the similarity reduction procedure. In this section,
some particular choices for the system (\ref{1.8}) will be considered. We
shall find equations describing concrete dynamical systems which admit
reduction through the similarity procedure to ordinary wave type equations
of the form (\ref{0.2}) and (\ref{0.3}). For the moment, we restrict the
forms (\ref{1.7}) of the infinitesimals $\varphi (t),\xi (x,t),\eta (x,t,u)$
to the following separable expressions:%
\begin{equation}
\text{ }\varphi =\varphi (t),\text{ }\xi =\xi (x,t)=\xi _{1}(x)\xi _{2}(t),%
\text{ }\eta =M(x,t)u=M_{1}(x)M_{2}(t)u\text{ }  \label{1.9}
\end{equation}%
The Lie operator (\ref{1.3}) becomes:%
\begin{equation}
U(x,t,u)=\varphi (t)\frac{\partial }{\partial t}+\xi _{1}(x)\xi _{2}(t)\frac{%
\partial }{\partial x}+M_{1}(x)M_{2}(t)u\frac{\partial }{\partial u}\text{ }
\label{1.10}
\end{equation}%
The general expressions of the invariants could be found if we should
consider the characteristic equations associated with the new generator (\ref%
{1.10}). These equations are:%
\begin{equation}
\frac{dt}{\varphi (t)}=\frac{dx}{\xi _{1}(x)\xi _{2}(t)}=\frac{du}{%
M_{1}(x)M_{2}(t)u}  \label{1.11}
\end{equation}%
By integrating the previous equations, two invariants are obtained with the
following expressions:%
\begin{equation}
I_{1}=\exp \left( \dint \frac{1}{\xi _{1}(x)}dx-\dint \frac{\xi _{2}(t)}{%
\varphi (t)}dt\right) ,\text{ }I_{2}=u\exp \left( -\frac{M_{2}(t)}{\xi
_{2}(t)}\dint \frac{M_{1}(x)}{\xi _{1}(x)}dx\right)  \label{1.12}
\end{equation}%
In the similarity reduction procedure two similarity variables have to be
considered:%
\begin{equation}
I_{1}=z,\text{ }I_{2}=\phi (z)  \label{1.13}
\end{equation}%
The invariants (\ref{1.13}) allow us, by an appropriate change of
coordinates $\{u,x,t\}\rightarrow \{z,\phi (z)\},$ to reduce the initial $%
(1+1)$ dimensional equation (\ref{1.1}) to an ordinary differential equation
of the form: 
\begin{equation}
\Omega ^{\prime }[z,\phi (z),\dot{\phi}(z),...]=0  \label{1.131}
\end{equation}

\subsection{Homogeneous wave type equation}

Our aim is now to select from the general dynamical systems described by (%
\ref{1.1}) the class of \ differential equations for which the equation (\ref%
{1.131}) can be reduced at a wave type equation of the form (\ref{0.2}):%
\begin{equation}
\frac{d^{2}\phi (z)}{dz^{2}}=0\Leftrightarrow \phi (z)=az+b  \label{1.14}
\end{equation}%
where $a$ and $b$ are arbitrary constants.

The previous solution, written in terms of the initial variable $(x,t)$,
leads to the following form of the solution $u(x,t)$ \ of (\ref{1.1}):%
\begin{equation}
u(x,t)=\left[ a\exp \left( \dint \frac{1}{\xi _{1}(x)}dx-\dint \frac{\xi
_{2}(t)}{\varphi (t)}dt\right) +b\right] \exp \left( \frac{M_{2}(t)}{\xi
_{2}(t)}\dint \frac{M_{1}(x)}{\xi _{1}(x)}dx\right)  \label{1.15}
\end{equation}%
For convenience reasons, we shall impose the following relations to be valid:%
\begin{equation}
\frac{\xi _{2}(t)}{\varphi (t)}=q,\text{ }\frac{M_{2}(t)}{\xi _{2}(t)}=v,%
\text{ }\dint \frac{1}{\xi _{1}(x)}dx\equiv P(x),\dint \frac{M_{1}(x)}{\xi
_{1}(x)}dx\equiv R(x)  \label{1.17}
\end{equation}%
with $q,v$ arbitrary constants.

In terms of notations (\ref{1.17}), the infinitesimals (\ref{1.9}) and the
solution (\ref{1.15}) become:%
\begin{eqnarray}
\varphi &=&\varphi (t),\text{ }\xi =q\frac{\varphi (t)}{\dot{P}(x)},\text{ }%
\eta =qv\frac{\varphi (t)\dot{R}(x)}{\dot{P}(x)}u\text{ }  \label{1.18} \\
u(x,t) &=&\left[ a\exp \left( P(x)-qt\right) +b\right] \exp \left(
vR(x)\right)  \label{1.19}
\end{eqnarray}%
The solution (\ref{1.19}) must verify the equation (\ref{1.1}) which
describes the analyzed model. This condition generates a differential system
of the form:%
\begin{eqnarray}
0 &=&q+2vA(x,t)\dot{R}(x)\dot{P}(x)+v^{2}A(x,t)[\dot{R}(x)]^{2}+A(x,t)\ddot{P%
}(x)+A(x,t)[\dot{P}(x)]^{2}+  \notag \\
&&+vA(x,t)\ddot{R}(x)+B(x,t)\dot{P}(x)+vB(x,t)\dot{R}(x)+C(x.t)  \notag \\
0 &=&v^{2}A(x,t)[\dot{R}(x)]^{2}+vA(x,t)\ddot{R}(x)+vB(x,t)\dot{R}(x)+C(x,t)
\label{1.20}
\end{eqnarray}

For an unitary analysis, it is necessary to describe the differential system
(\ref{1.8}), obtained in the previous subsection, in terms of the functions $%
P(x)$ and $R(x)$ introduced by (\ref{1.17}). Using the expressions (\ref%
{1.18}) we obtain the following differential system:%
\begin{equation*}
\text{ }\varphi A_{t}P_{x}^{2}+q\varphi A_{x}P_{x}+\varphi
_{t}AP_{x}^{2}+2q\varphi AP_{2x}=0
\end{equation*}%
\begin{equation*}
\varphi B_{t}P_{x}^{4}+q\varphi B_{x}P_{x}^{3}+q\varphi
BP_{2x}P_{x}^{2}+q\varphi _{t}P_{x}^{3}+\varphi _{t}BP_{x}^{4}+
\end{equation*}%
\begin{equation}
+2vq\varphi AR_{2x}P_{x}^{3}-2vq\varphi AR_{x}P_{2x}P_{x}^{2}+q\varphi
AP_{3x}P_{x}^{2}-2q\varphi AP_{x}P_{2x}^{2}=0  \label{1.210}
\end{equation}%
\begin{equation*}
\varphi C_{t}P_{x}^{4}+q\varphi C_{x}P_{x}^{3}+qv\varphi
BR_{2x}P_{x}^{3}-qv\varphi BR_{x}P_{2x}P_{x}^{2}+\varphi
_{t}CP_{x}^{4}+vq\varphi AR_{3x}P_{x}^{3}-
\end{equation*}%
\begin{equation*}
+vq\varphi AR_{x}P_{3x}P_{x}^{2}-2vq\varphi
AR_{2x}P_{2x}P_{x}^{2}++2qv\varphi AR_{x}P_{x}P_{2x}^{2}-qv\varphi
_{t}R_{x}P_{x}^{3}=0
\end{equation*}

\textbf{Conclusion: }Our problem is to find the class of \ $(1+1)$
evolutionary equations of type (\ref{1.1}) which could be reduced by the
similarity approach to an ordinary wave type equation. Solving this problem
is equivalent with searching the solutions of the system described by
equations (\ref{1.20}) and (\ref{1.210}).

\textbf{Remark 1:} The system (\ref{1.20})-(\ref{1.210}) can be solved
following two paths: (\textit{i}) by choosing a concrete dynamical system,
that is to say concrete expressions for the functions $A(x,t),$ $B(x,t),$ $%
C(x,t)$ and trying to find out if this equation admits or not solution of
the type (\ref{1.19}). Now the unknown functions of the system are $\varphi
(t),$ $P(x),$ $R(x)$ defined by (\ref{1.17}); (\textit{ii}) by considering $%
A(x,t),$ $B(x,t),$ $C(x,t)$ as unknown functions and by choosing $\varphi
(t),$ $P(x),$ $R(x).$ This is the way we shall follow in the next section.

\textbf{Remark 2: }In the case (\textit{ii}) the general solutions obtained
by computational way can be expressed as:%
\begin{eqnarray}
A(x,t) &=&F\left( \frac{qt-P(x)}{q}\right) \exp \left[ G(x)\right]  \notag \\
B(x,t) &=&\frac{\left[ -2F\left( \frac{qt-P(x)}{q}\right) \left( \frac{1}{2}[%
\dot{P}(x)]^{2}+v\dot{R}(x)\dot{P}(x)+\frac{1}{2}\ddot{P}(x)\right) \exp %
\left[ -G(x)\right] -q\right] }{\dot{P}(x)}  \notag \\
C(x,t) &=&\frac{v\left[ F\left( \frac{qt-P(x)}{q}\right) \left[ \dot{R}(x)%
\ddot{P}(x)+\dot{P}(x)(-\ddot{R}(x)+\dot{R}(x)(v\dot{R}(x)+\dot{P}(x)))%
\right] \exp ^{\left[ -G(x)\right] }+q\dot{R}(x)\right] }{\dot{P}(x)}  \notag
\\
&&  \label{1.22}
\end{eqnarray}%
where 
\begin{equation}
G(x)=-\int^{x}\frac{2(D^{(2)})(P)(a)\varphi \left( \frac{P(a)+qt-P(x)}{q}%
\right) q+\left[ D(P)(a)\right] ^{2}D(\varphi )\left( \frac{P(a)+qt-P(x)}{q}%
\right) }{D(P)(a)\varphi \left( \frac{P(a)+qt-P(x)}{q}\right) q}da
\label{1.22'}
\end{equation}%
These solutions are valid for arbitrary constants $q,v$ and for an arbitrary
function $F\left( \frac{qt-P(x)}{q}\right) .$

\subsection{Harmonic oscillators}

Let us consider now that the similarity reduction equation is an ordinary
oscilator type equation of the form (\ref{0.3}):

\begin{equation}
\frac{d^{2}\phi (z)}{dz^{2}}+k^{2}\phi (z)=0  \label{21.13}
\end{equation}%
It has the solution:%
\begin{equation}
\phi (z)=a\sin (kz)+b\cos (kz)  \label{21.14}
\end{equation}%
Here $k,$ $a$ and $b$ are arbitrary constants.

To writte down the previous solution in terms of the initial variable $%
u(x,t) $ \ means that the solution of (\ref{1.1}) should have the form:%
\begin{eqnarray}
u(x,t) &=&a\exp \left( \frac{M_{2}(t)}{\xi _{2}(t)}\dint \frac{M_{1}(x)}{\xi
_{1}(x)}dx\right) \sin \left( \sqrt{k}\exp \left( \dint \frac{1}{\xi _{1}(x)}%
dx-\dint \frac{\xi _{2}(t)}{\varphi (t)}dt\right) \right) +  \label{21.15} \\
&&+b\exp \left( \frac{M_{2}(t)}{\xi _{2}(t)}\dint \frac{M_{1}(x)}{\xi _{1}(x)%
}dx\right) \cos \left( \sqrt{k}\exp \left( \dint \frac{1}{\xi _{1}(x)}%
dx-\dint \frac{\xi _{2}(t)}{\varphi (t)}dt\right) \right)  \notag \\
&&  \notag
\end{eqnarray}

For convenience reasons, we shall impose again the following relations to be
valid:%
\begin{equation}
\frac{\xi _{2}(t)}{\varphi (t)}=q,\text{ }\frac{M_{2}(t)}{\xi _{2}(t)}=v,%
\text{ }\dint \frac{1}{\xi _{1}(x)}dx\equiv P(x),\dint \frac{M_{1}(x)}{\xi
_{1}(x)}dx\equiv R(x)  \label{21.17}
\end{equation}%
with $q,v$ arbitrary constants.

In terms of notations (\ref{21.17}), the infinitesimals (\ref{1.9}) and the
solution (\ref{21.15}) become:%
\begin{equation}
\varphi =\varphi (t),\text{ }\xi =q\frac{\varphi (t)}{\dot{P}(x)},\text{ }%
\eta =qv\frac{\varphi (t)\dot{R}(x)}{\dot{P}(x)}u\text{ }  \label{21.18}
\end{equation}%
\begin{eqnarray}
u(x,t) &=&\left[ a\sin (k\exp \left( P(x)-qt\right) )+b\cos (k\exp \left(
P(x)-qt\right) )\right] \exp \left( vR(x)\right)  \notag \\
&&  \label{21.19}
\end{eqnarray}%
The solution (\ref{21.19}) must verify the equation (\ref{1.1}) which
describes the analyzed model. This condition generates the vanishing of the
coefficient function $A(x,t)$ and two other differential equations of the
form:%
\begin{eqnarray}
A(x,t) &=&0  \notag \\
q+B(x,t)\dot{P}(x) &=&0  \label{21.20} \\
vB(x,t)\dot{R}(x)+C(x,t) &=&0  \notag
\end{eqnarray}

For an unitary analysis, it is again necessary to describe the general
differential system (\ref{1.8}) obtained in the previous section, in terms
of the functions $P(x)$ and $R(x)$ introduced by (\ref{21.17}). Taking into
account the equations (\ref{21.18}), we obtain the following differential
system:%
\begin{eqnarray}
0 &=&\varphi B_{t}P_{x}^{4}+q\varphi B_{x}P_{x}^{3}+q\varphi
BP_{2x}P_{x}^{2}+q\varphi _{t}P_{x}^{3}+\varphi _{t}BP_{x}^{4}  \notag \\
0 &=&\varphi C_{t}P_{x}^{4}+q\varphi C_{x}P_{x}^{3}+qv\varphi
BR_{2x}P_{x}^{3}-qv\varphi BR_{x}P_{2x}P_{x}^{2}+  \notag \\
&&+\varphi _{t}CP_{x}^{4}-qv\varphi _{t}R_{x}P_{x}^{3}  \label{21.21}
\end{eqnarray}%
The system (\ref{21.20})-(\ref{21.21}) can be solved following two paths: (%
\textit{i}) by choosing a concrete dynamical system, that is to say concrete
expressions for the functions $B(x,t),$ $C(x,t)$ and trying to find out if
this equation admits or not solution of the type (\ref{21.19}). Now the
unknown functions of the system are $\varphi (t),$ $P(x),$ $R(x)$ defined by
(\ref{21.17}); (\textit{ii}) by considering $B(x,t),$ $C(x,t)$ as unknown
functions and by choosing $\varphi (t),$ $P(x),$ $R(x).$

This second case is the way we are interested in to follow and, in this
case, the general solutions obtained by computational way can be expressed
as:%
\begin{equation}
B(x,t)=\frac{-q}{\dot{P}(x)},\text{ }C(x,t)=\frac{vq\dot{R}(x)}{\dot{P}(x)}
\label{21.22}
\end{equation}%
or in terms of the coefficient functions $\varphi (t),$ $\xi (x,t),$ $M(x,t)$
which appear in the general Lie symmetry operator (\ref{21.10}), in the
equivalent forms:%
\begin{equation*}
B(x,t)=\frac{-\xi (x,t)}{\varphi (t)},\text{ }C(x,t)=\frac{M(x,t)}{\varphi
(t)}
\end{equation*}

\subsection{Rossby type symmetries}

The equation for coupled gravity, inertial and Rossby waves in a rotating,
stratified atmosphere using the $\beta $-plane approximation (which
simplifies the spherical geometry whilst retaining the essential dynamics)
and the Boussinesq approximation which filters out higher frequency acoustic
waves can be written in ($2+1)-$dimensions in the form \cite{McKenzie}:

\begin{equation}
\frac{\partial }{\partial t}\left( \frac{\partial ^{2}}{\partial x^{2}}+%
\frac{\partial ^{2}}{\partial y^{2}}\right) u_{y}=-\beta \frac{\partial u_{y}%
}{\partial x}  \label{21.23}
\end{equation}%
As we mentioned, this equation describes the coupling between the inertial,
the gravity and the Rossby waves, but also the shallow water in an ocean of
depth H. It was proven \cite{Rossby1} that reducing the model to ($1+1)-$%
dimensions, $(x,t)$, it admits some very simple Lie symmetries of the form:

\begin{equation}
\varphi (t)=ct+c_{1},\text{ }\xi (x,t)=cx+f(t),\text{ }\eta (u)=-3cu
\label{21.24}
\end{equation}%
with $f(t)$ arbitrary function and $c_{0},$ $c$ arbitrary constants. Despite
this fact, it is still difficult to find explicite solutions for the
equation (\ref{21.23}). This is why, we shall consider another approach: we
shall impose the Lie symmetries (\ref{21.24}) to our general equation (\ref%
{1.1}) and we shall try to find the class of equations which observe them.
This means that we have in fact to impose the Rossby symmetries (\ref{21.24}%
) to the system (\ref{1.8}). It will take the form:

\begin{eqnarray*}
(ct+c_{1})A_{t}+(cx+c_{2})A_{x}+3cA &=&0 \\
-(ct+c_{1})B_{t}-(cx+c_{2})A_{x}-2cB &=&0 \\
-(ct+c_{1})C_{t}-(cx+c_{2})C_{x}-cC &=&0
\end{eqnarray*}%
with the unknown functions $A(x,t),$ $B(x,t),$ $C(x,t).$

This system admits the solutions:%
\begin{eqnarray*}
A(x,t) &=&\frac{F\left( x(ct+c_{1})-c_{2}t\right) }{(ct+c_{1})^{3}}=\frac{%
F\left( x\varphi (t)-c_{2}t\right) }{\left[ \varphi (t)\right] ^{3}} \\
B(x,t) &=&\frac{G\left( x(ct+c_{1})-c_{2}t\right) }{(ct+c_{1})^{2}}=\frac{%
G\left( x\varphi (t)-c_{2}t\right) }{\left[ \varphi (t)\right] ^{2}} \\
C(x,t) &=&\frac{H\left( x(ct+c_{1})-c_{2}t\right) }{ct+c_{1}}=\frac{H\left(
x\varphi (t)-c_{2}t\right) }{\varphi (t)}
\end{eqnarray*}%
with $F,$ $G,$ $H$ arbitrary functions of their arguments. These expressions
give us equations of the form (\ref{1.1}) which are equivalent from the
point of view of their symmetries with the Rossby equation.

\section{Conclusions}

The problem of finding exact solutions for nonlinear differential equations
plays an important role in the study of nonlinear dynamics. There are many
ways of tackling with it. One of them is based on the Lie symmetry method.
This method supposes to find the symmetries of the system and, on this
basis, to try to determine the general or some particular solutions of the
equations. There is a direct approach in which the symmetries of a given
equation are obtained, but also an inverse problem has been formulated \cite%
{Cimpoiasu}. A step forward for this latter approach is represented by the
use of similarity reduction, a procedure which allows the reduction of the
number of degrees of freedom and, by that, simplifies the problem of solving
the equation. This paper used this approach and determined a class of $(1+1)$
dimensional second order differential equations which can be reduced to
ordinary wave-type equations with simple solutions. Using the Lie symmetry
and the similarity reduction procedures, some particular cases of the
equation (\ref{1.1}) arise as good candidates of equations which could be
used as generalization of the linear wave type equations describing complex
atmospheric phenomena. Moreover, following our method, we were able to write
down the solutions of these equations, solutions which otherwise could be
derived by computational methods, but in a very complicated form. Another
interesting results of our paper consisted in the fact that a complicated,
nonintegrable equation, the Rossby equation, could be replaced by another,
simpler equation, which have similar symmetries. The paper is important both
by these results, but also as a methodological approach in finding exact
solutions through similarity reduction procedure. We have shown how,
starting from a particular form of solution for the reduced equation we
could recover the solution of \ a most complicated problem, defined in a
space with more than one dimensions. We tackled out a particular case,
looking only for linear solutions of the reduced equation and considering
that the coefficient functions appearing in the symmetry operators are
separable. The problem can be extended for other cases, too.

\textbf{Acknowledgements}: This work was supported by CNCSIS -UEFISCSU,
project number PNII - IDEI code 418/2008.

\bigskip

\end{document}